\documentclass[aps, prl, twocolumn, nofootinbib]{revtex4-1} 
\pdfoutput=1
\usepackage{amsmath}
\usepackage{amssymb}
\usepackage{braket}

\usepackage{graphicx}

\begin{document}

\title{Learning the {E}instein-{P}odolsky-{R}osen correlations on a {R}estricted 
{B}oltzmann {M}achine}

\author{Steven Weinstein}
\affiliation{Perimeter Institute for Theoretical Physics, 31 Caroline St. N., 
Waterloo, ON N2L 2Y5, Canada}
\email{sweinstein@pitp.ca}
\affiliation{Department of Physics and Department of Philosophy, University of 
Waterloo, 200 University Ave W., Waterloo, ON N2L 3G1, Canada}
\email{sw@uwaterloo.ca}



\begin{abstract}
We construct a hidden variable model for the EPR correlations using a 
Restricted Boltzmann Machine.  The model reproduces the expected 
correlations and thus violates the Bell inequality, as required by Bell's theorem. 
Unlike most hidden-variable models, this model does not violate the 
\emph{locality} assumption in Bell's argument. Rather, it violates 
\emph{measurement independence}, albeit in a decidedly non-conspiratorial 
way.
\end{abstract}

\maketitle

\section{Introduction}
The Restricted Boltzmann Machine (RBM) is a machine learning model dating to 
the early 1980s \cite{Smo86} which has enjoyed renewed interest over the last 
decade for its utility in deep learning \cite{Hin14}. Inspired by physical models of 
locally interacting spins, Boltzmann machines are able to ``learn" underlying 
patterns in data sets by systematically adjusting their weights in such a way that 
the equilibrium state of the network, a Boltzmann distribution over the network 
configurations, expresses the structural correlations in the data set. In this paper, 
we show that one can train a simple RBM to model the data from an EPR 
(Einstein-Podolsky-Rosen) experiment \cite{EPR35}. The hidden units of the 
machine correspond to a set of hidden variables, giving us a \emph{local}, 
stochastic hidden variable model for the puzzling correlations seen in the 
experiment. (See also \cite{Ver13} for an earlier Ising-inspired model.)

First, we provide a brief look at the essentials of Boltzmann machines and the 
subset of these called Restricted Boltzmann Machines. Second, we review Bell's 
theorem, which shows that any model of the EPR correlations having the 
property of \emph{measurement independence} (also known as \emph{statistical 
independence}) must be nonlocal. Third, we describe an RBM that generates 
EPR data and thereby provides a model of the EPR correlations. Finally, we 
discuss the essentially local nature of the model, and show the rather natural, 
uncontrived way in which measurement independence fails, yielding a local 
theory that violates the Bell inequality and reproduces the predictions of 
quantum mechanics. 

\section{Boltzmann Machines}

Boltzmann machines \cite{HinSej83, AckHinSej85} are stochastic models 
inspired by Hopfield networks \cite{Hop82} and more generally Ising-type 
models, models of interacting two-state spin systems at a finite temperature. 
They provide a model of collective computation, one that may shed light on the 
way in which the brain processes information, since neurons can be idealized as 
simple two-state (on/off) systems or ``units'' massively interconnected with other 
similar units.  

\begin{figure}[h]
  \begin{minipage}[h]{0.45\linewidth}
    \includegraphics[width=.9\textwidth]{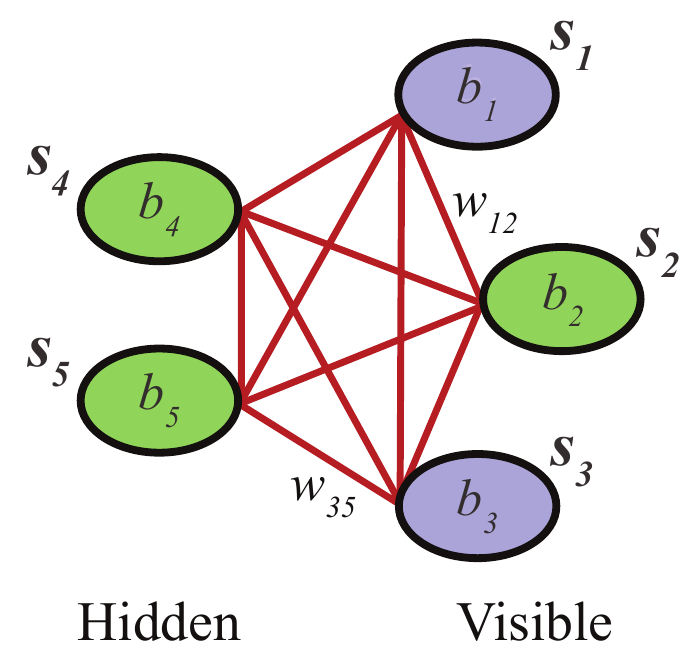}
    \caption{General Boltzmann machine with 3 visible and 2 hidden units.}
    \label{fig:Boltz}
  \end{minipage}
   \hspace{0.5cm}
  \begin{minipage}[h]{0.45\linewidth}
    \includegraphics[width=.9\textwidth]{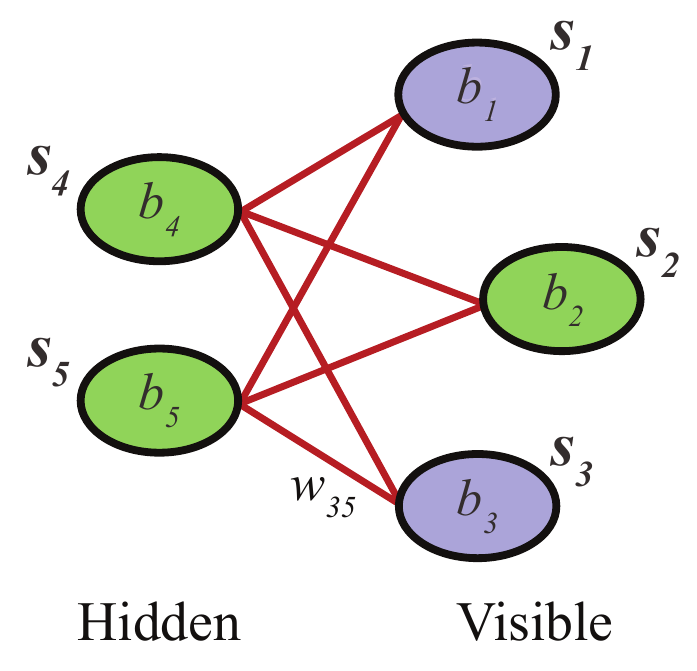}
    \caption{Restricted Boltzmann Machine. Green is \emph{on}, blue is 
\emph{off}.}
    \label{fig:RBM1}
  \end{minipage}
\end{figure}

General Boltzmann machines can have an arbitrarily large number of binary 
units, connected in any topology one likes (see Figure \ref{fig:Boltz}). 
Once the topology is fixed, what distinguishes one instantiation from another are 
the weights of the links connecting the units (analogous to the interaction 
strengths in an Ising model) and the biases of the individual units (analogous to 
the coupling to an inhomogeneous external field). The idea behind a Boltzmann 
machine is that, given a function to define the energy of the network -- a 
Hamiltonian -- and an appropriate stochastic dynamics for updating the (binary) 
states of the units, the network will have a natural equilibrium distribution of 
configurations given by the Boltzmann distribution, and these distributions can 
thereby represent probability distributions over sets of hypotheses, states of the 
world, or what have you.

The total energy of a Boltzmann machine is the sum of the self-energies of each 
unit and the interaction energies between neighboring units:
\begin{equation}\label{hamiltonian}
E = -\sum_i b_i \, s_i  - \sum_{i<j} w_{ij} \, s_i \, s_j .
\end{equation}
(We write $i < j$ in order to avoid double-counting the interaction energy 
between pairs of units.) In machine learning, the coefficients $b_i$ are called 
\emph{biases}, the interaction strengths $w_{ij}$ are called \emph{weights}, and 
the states $s_i$ take values 1 or 0, rather than $+1$ or $-1$ as is conventional 
in the Ising model.

The dynamics of a Boltzmann machine are stochastic, as in the Ising model at 
finite temperature. The probability that a given unit $i$ will turn (or remain) on is 
a function of the difference in the total network energy resulting from the unit's 
being on and off:
\begin{equation}
\begin{split}
{\Delta}E_{i} &= E_{s_i=0} - E_{s_i=1}\\
&= {b_i} +\sum_j w_{j}\,{s_j} .
\end{split}
\end{equation}
Thus the difference in energy ${\Delta}E_{i}$ between the two possible states of 
$i$ depends on the weights of the connections to other units, and on the bias of 
unit $i$. The update rule is
\begin{equation}\label{updaterule}
P(s_i = 1) = \frac{1}{1 + e^{-{\Delta}E_{i}}} .
\end{equation}
(The temperature does not appear in this expression because we train the 
machine at fixed temperature, such that $kT=1$). Updates are generally done in 
an asynchronous manner.

The state of the entire network at a given moment is given by a vector $\bold{s}
$. Given the energy function (Eq. \ref{hamiltonian}) and update rule (Eq. 
\ref{updaterule}), the probability that the network will be in configuration $\bold{s}
$ is given by the Boltzmann distribution
\begin{equation}
\label{boltzmanndistribution}
P({\bold{s}} ) = \frac{e^{-E(\bold{s})}}{\sum_k {e^{-E({\bold{s_k}})}}},
\end{equation}
where the index $k$ ranges over all possible states of the network.

For most purposes, we make a nominal distinction between visible and hidden 
units, so that $\bold{s} = (\bold{v}, \bold{h})$. The visible units represent 
observed (or observable) properties of the objects of interest, and the relative 
frequencies of 1s and 0s on these units encode the correlations in the world we 
are interested in. The hidden units encode the structural properties behind these 
correlations, structure that the machine learns so as to be able to generate and 
predict the observed properties.\footnote{Goodfellow \emph{et. al.} \cite{GSS14} 
give examples that show that this interpretation of the significance of the hidden 
units should be taken with at least a grain of salt.} Training (or ``learning," in the 
parlance of the field) a Boltzmann machine is a procedure to get it to reproduce 
the correlations in the data as correlations on the visible units. It involves picking 
a more-or-less random set of weights and biases, checking the output (the 
frequencies of the various visible configurations), and making adjustments to the 
weights and biases so that the output approaches the target, i.e. so that it 
approximates the data in the training distribution. 

Efficient training of large Boltzmann machines with unrestricted connections 
between the units is highly non-trivial in part because the partition function (the 
denominator of Eq. (\ref{boltzmanndistribution})) is difficult to calculate, and in 
part because the visible units depend not only on the hidden units but on the 
other visible units. We modeled the EPR correlations using a Restricted 
Boltzmann Machine (RBM), a particular kind of Boltzmann machine in which the 
$m$ visible units and $n$ hidden units form two layers, with no intra-layer 
connections (see Figure \ref{fig:RBM1}). This is a bipartite, undirected graph, 
and the energy function (Eq. \ref{hamiltonian}) above takes the form 
\begin{equation}\label{RBMhamiltonian}
E(\bold{v},\bold{h}) = -\bigg( \sum_{i=1}^m c_i v_i  + \sum_{j=1}^n d_j h_j + 
\sum_{i=1}^m\sum_{j=1}^n w_{ij} v_i h_j \bigg)
\end{equation}
where $c_i$ and $d_j$ are the biases for the visible and hidden units, 
respectively. 

The functional form of Eq. (\ref{RBMhamiltonian}) implies that the units within 
each layer are conditionally independent.  This makes the machine considerably 
easier to train than a general Boltzmann machine \cite{Smo86, Hin02}. The 
conditional probabilities $P(\bold{v} | \bold{h})$ and $P(\bold{h} | \bold{v})$ take 
the simple product form
\begin{align}
\label{CIa}
P(\bold{v} | \bold{h}) &= \prod_{i=1}^m P(v_i | \bold{h})\\
\label{CIb}
P(\bold{h} | \bold{v}) &= \prod_{j=1}^n P(h_j|\bold{v}) . 
\end{align}
The fact that the conditional probabilities factor in this way not only makes 
learning easier, but it also allows us to retain the idea that the detector settings 
in the EPR experiment can be regarded as independent degrees of freedom, 
which may thus be freely chosen.

\section{EPR and Bell's Theorem}

Bell's theorem is a demonstration that the predictions of quantum mechanics for 
the EPR experiment are incompatible with two assumptions, the most prominent 
of which is a particular kind of \emph{locality}, sometimes called Bell-locality or 
strong locality \cite{Bell64}.

In its modern guise due to Bohm \cite{Boh51}, the EPR experiment involves spin 
measurements on a pair of particles that have been prepared in a maximally 
entangled state. There are two stations, which we will refer to as $A$ and $B$, 
and detectors with two possible settings at each station, the different settings 
corresponding to measurements of different components of spin. 
The detector settings $\alpha = \{a, a'\}$ and $\beta = \{b, b'\}$ and 
measurement outcomes $x_\alpha = \{+1, -1\}$ and $x_\beta = \{+1, -1\}$ are 
two-valued, so we can treat them as Bernoulli random variables. A theory or 
model of the experiment consists of one or more ``states'' $\lambda \in 
\Lambda$, each of which implies a joint probability distribution over $\alpha$, $
\beta$, $x_\alpha$, and $x_\beta$.

Bell's locality criterion is intended to encode the assumption that the underlying 
theory that accounts for the correlations will be local in the sense that the 
outcome $x_{\alpha}$ at A is independent of both the detector setting $\beta$ at 
B (``parameter independence'') and the outcome $x_{\beta}$ at B (``outcome 
independence''). This is equivalent to assuming that the joint probability 
distribution is equal to the product of the two marginal distributions, and is 
therefore often referred to as factorizability:
\begin{equation}\label{locality}
P(x_\alpha, x_\beta |  \alpha, \beta, \lambda) = P(x_\alpha | \alpha, \lambda)  P 
(x_\beta | \beta, \lambda). 
\end{equation}

There is an additional, often tacit assumption called \emph{measurement 
independence}, which we will have occasion to discuss further. In brief, it is the 
assumption that the state $\lambda$ is independent of the detector settings $
\alpha$ and $\beta$, so that
\begin{equation}\label{MI}
 P(\lambda | \alpha, \beta) = P(\lambda).
\end{equation}

Bell showed that the assumption of locality, plus the assumption of 
measurement independence, imply that the correlations $C(\alpha,\beta)$ 
between measurement results at various detector settings should satisfy an 
inequality known as the Bell-inequality, later generalized to the CHSH-Bell 
inequality \cite{CHSH69}:
\begin{equation}\label{CHSH}
S = |C(a, b) + C(a, b') + C(a', b) - C(a', b')| \leq 2 .
\end{equation}
For appropriate choices of the detector angles $a$, $a'$, $b$, and $b'$, this 
inequality is violated by quantum mechanics. If one takes the singlet state 
\begin{equation}
\psi = \frac{1}{\sqrt{2}}\big(\ket{+-} - \ket{-+}\big)
\end{equation}
as $\lambda$ and chooses $a=0$, $a'=\pi/2$, $b= \pi/4$, and $b'=-\pi/4$ radians 
as orientations for the measuring apparatuses, quantum mechanics predicts that 
the correlation coefficients have values such that $S = 2\sqrt{2} = 2.828$, which 
violates the inequality (see the \emph{Theory} column of Table \ref{tab:corr}).
\begin{table}[h]
\begin{tabular}{l | r r r r}
\hline
 & Theory & Data & Model &\\
 \hline
$C(a,b)$ & $-0.707$ & $-0.713$ & $ -0.711$ &\\
$C(a,b')$ & $-0.707$ &$-0.701$ & $ -0.699$ &\\
$C(a',b)$ & $-0.707$ &$-0.714$ & $ -0.713$ &\\
$C(a',b')$ & $0.707$ &$0.709$ & $ 0.704$&\\
\hline
\end{tabular}
\caption{\label{tab:corr}Correlation coefficients for detector settings $a=0$, 
$a'=\pi/2$, $b= \pi/4$, and $b'=-\pi/4$.}
\end{table}

As the quantum-mechanical predictions are borne out by experiment \cite{FC72, 
Aspect81, Aspect82}, \emph{any} theory that accounts for the experimental data 
must violate at least one of the two assumptions used to derive the inequality.  
Quantum mechanics, as it happens, violates locality, as do Bohmian mechanics 
\cite{Bohm52a,Bohm52b}, Nelson's stochastic mechanics 
\cite{Nel66,Nel85,Smo06}, and the GRW spontaneous collapse model 
\cite{GRW86, GPR90}. 

The RBM model does not violate locality; rather, it violates measurement 
independence. Models of this sort are far less common, as they have seemed to 
many (including Bell \cite{Bell75, Bell90b}, Shimony \emph{et al.} \cite{BSHC85} 
and others) to preclude ``free will'' on the part of the experimenter, or to involve 
some sort of conspiracy or fine-tuning on the part of nature \cite{WS15}. Many of 
these models are billed as retro-causal \cite{Cram86, ACE15, PW17, LP17}, but 
there are others, sometimes called ``superdeterministic'' \cite{Brans88, Hall10, 
BG11, Hall16}, at least one of which \cite{Hooft03, Hooft14} invokes a nonlocal, 
spacelike constraint of the sort suggested in Weinstein \cite{SW09}. The RBM 
model, we will see, is different again.

\section{Training the EPR Machine}
Training a Boltzmann machine, restricted or otherwise, involves confronting it 
with data and getting it to revise its model of the data in such a way as to 
reproduce the correlations in the data as correlations on the visible units. 

In our case, the data include both the outcomes of experiments and the detector 
settings, since we are interested in exhibiting correlations between detector 
settings and pairs of outcomes. Rather than use data from actual runs of an 
EPR-type experiment to train our machine, we simulated 100,000 runs using a 
simple Python script, deriving outcomes for randomly chosen detector settings 
using ordinary quantum mechanical calculations.  The resulting correlation 
coefficients are in the \emph{Data} column of Table \ref{tab:corr}.  

\begin{figure}[h]
\begin{center}
\includegraphics[width=0.9\linewidth]{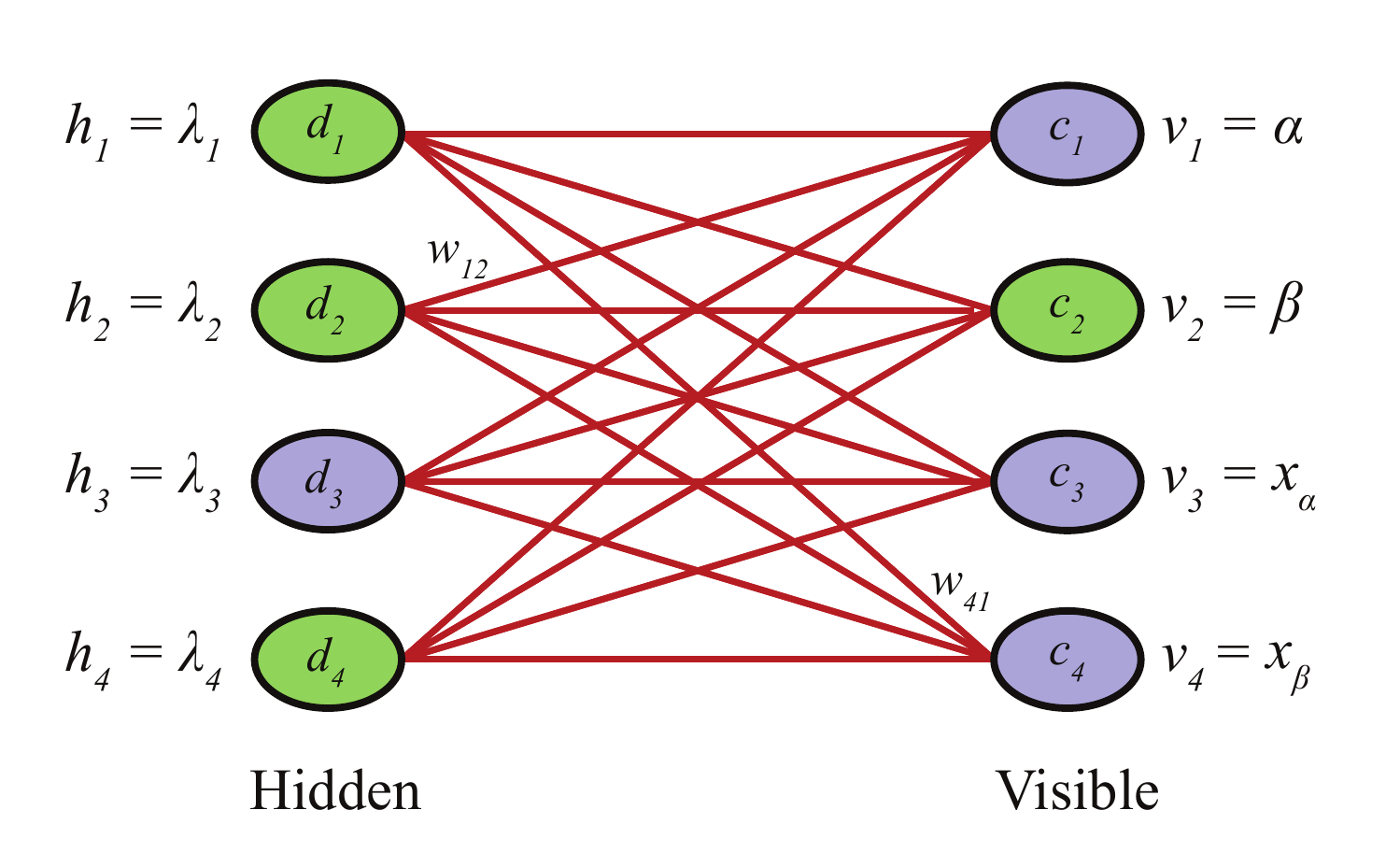}
\caption{RBM for EPR: Green is \emph{on} (0); blue is \emph{off} (1). In the 
configuration shown, $\bold{\lambda}= (1,1,0,1)$, $\alpha = a'$, $\beta = b$, and 
$x_\alpha=x_\beta=-1$.}
\label{fig:RBMcrop}
\end{center}
\end{figure}

The RBM we constructed has four visible and four hidden units (see Figure 
\ref{fig:RBMcrop}). 
The units $v_1$ and $v_2$ represent the detector settings $\alpha$ and $
\beta$, respectively, while $v_3$ and $v_4$ represent $x_\alpha$ and 
$x_\beta$, the outcomes of measurements at those detectors.  For example, 
consider the correlation between the outcomes with settings $a$ and $b$.  The 
observed value of $C(a,b)$ in our training data was $-0.713$, which means that 
when the detectors were set to measure $a$ and $b$, the results were different 
(perfectly anti-correlated) around 85.7\% of the time and the same (perfectly 
correlated) around 14.3\% of the time.  The goal is to reflect this as a correlation 
between the on/off probabilities of the visible units $v_1, v_2, v_3, v_4$ such 
that
\begin{align}
P(v_3 = v_4 | (v_1,v_2) = (0, 0)) &\simeq 0.143 \\
P(v_3 \neq v_4 | (v_1,v_2) = (0, 0)) &\simeq 0.857 .
\end{align}
Training the machine involves initializing the network with random weights and 
biases, and adjusting them in an iterative process so as to bring the distribution 
on the visible units in line with the data.

Because of the restricted topology of the network, the rule for adjusting the 
weights is both simple and local, despite the extensive interconnection of the 
units. Recall that we are trying to get the visible layer to align with the data, so 
we are especially interested in the $\bold{v} = (v_1, v_2, v_3, v_4)$ portion of 
the overall configuration $(\bold{v}, \bold{h})$. From Eq. (\ref{updaterule}) and 
Eq. (\ref{RBMhamiltonian}), it follows that
\begin{equation}
\frac{\partial{\log{p(\bf{v})}}}{\partial{w_{ij}}} = \langle v_i h_j\rangle{_{data}} - 
\langle v_i h_j\rangle{_{model}} .
\end{equation}
As such, the weight update rule is of the form:
\begin{equation}
\Delta{w_{ij}}= \epsilon(\langle v_i h_j\rangle{_{data}} - \langle v_i 
h_j\rangle{_{model}}),
\end{equation}
where $\epsilon$ is a small, real-valued parameter colloquially known as the 
\emph{learning rate}. Note that the expectation value $\langle v_i h_j\rangle$ is 
simply the probability that both components will have the value 1, i.e., that they 
will both be on.

Because the layers are independent, we have
\begin{eqnarray}\label{condind}
P(h_j = 1 | \bold{v}) &= \sigma (d_j + \sum_i v_i w_{ij})\\
P(v_i = 1 | \bold{h}) &= \sigma (c_i + \sum_j h_j w_{ij}) ,
\end{eqnarray}
where $\sigma(x) = 1/(1 + e^{-x})$. The training data from the simulation gives 
us a distribution over visible vectors $\bold{v}$.  Each visible vector determines 
a probability for the value of each hidden unit, and so we are able to determine $
\langle v_i h_j\rangle{_{data}}$ in a straightforward fashion.   

The determination of $\langle v_i h_j\rangle{_{model}}$ must be approximated 
for RBMs with more than a small number of units, as one otherwise needs to 
calculate the probability of every configuration explicitly, since there are no data 
to condition over. The number of configurations grows exponentially with the 
number of units, so this gets out of hand quickly. However, methods exist for 
efficient approximation \cite{Hin12}, and our results were obtained in this 
manner. 

\section{Results}
Training the model on 100,000 trials using persistent contrastive divergence 
\cite{Tiel08} yielded a Restricted Boltzmann Machine with the weights and 
biases in Table \ref{tab:weightsbiases}.  
\begin{table}[h]
\begin{tabular}{r r r r r r}
\hline
  & \vline &$h_1$ & $h_2$ & $h_3$ & $h_4$ \\
  & \vline & $(-3.320)$ & $(-1.015)$ & $(-0.933)$ & $(-3.753)$\\
  \hline
$v_1$ & $(-5.026)\ $\vline   & $2.652$ & $3.527$ & $3.546$ & $-2.456$ \\
$v_2$ & $(-4.872)\ $\vline  & $-2.664$ & $3.575$ & $3.585$ & $2.471$ \\
$v_3$ & $(-3.467)\ $\vline  & $3.343$ & $-5.587$ & $5.578$ & $ 3.717$ \\
$v_4$ & $(-3.464)\ $\vline   & $3.326$ & $5.577$ & $-5.592$ & $3.721$ \\
\hline
\end{tabular}
\caption{\label{tab:weightsbiases} Weights $w_{ij}$ of the connections between 
visible and hidden units. Biases for the individual units in parentheses.}
\end{table}
The correlation coefficients for this model are given in the \emph{Model} column 
of Table \ref{tab:corr}. They are a remarkably close fit to the actual data. The 
model could be further refined, but there is no point, as it would merely overfit 
the data. We have thus successfully modeled the key properties of simulated 
EPR data as the output of a Restricted Boltzmann Machine with four hidden 
units.  

\section{Discussion}

The RBM model is a stochastic hidden variable theory of a very interesting sort. 
The binary states of the four hidden units are the hidden variables in the model, 
giving rise to  $2^4 = 16$ hidden states $\lambda$, each of which generates a 
distinct set of probabilities for the activations of the visible units. The Bell locality 
condition (Eq. \ref{locality}) is not violated, since the probabilities of the 
outcomes $x_\alpha$ at A and $x_\beta $ at B are independent of the detector 
settings and outcomes at B and A, respectively, in virtue of the conditional 
independence of the visible units (Eq. \ref{CIa}). This is a direct result of the 
topology of the RBM: there are no connections between visible units (or between 
hidden units). It has a kind of connectivity that would easily allow it to be 
embedded in a relativistic spacetime, with the hidden units and visible units each 
mutually spacelike, and each of the hidden units timelike or lightlike to each of 
the visible units. 

One might be forgiven for thinking of the RBM model as a retrocausal model, 
especially if one takes the topology of the model as suggesting a causal 
structure of this sort.  That is, one might think that there are boundary conditions 
in the past (the state preparation -- not modeled here) and the future (the 
detector settings), and that the future boundary conditions, which are after all 
freely specifiable, retroactively bring about changes in the prior state of the 
system.  

But the RBM model described here is not that. The weights and biases of the 
network are independent of the detector settings, and so is the probability 
distribution. Predictions are conditional probabilities, specifically probabilities for 
measurement outcomes conditioned on the states of the detectors. The model is 
atemporal and therefore acausal. There is no dynamics that propagates a 
change from the future to the past, or from the past to the future. Outcomes are 
generated not by the dynamical evolution of an initial condition, but in the way of 
equilibrium statistical mechanics. The dynamics lies in the update rule, Eq 
(\ref{updaterule}), which generates a stationary probability distribution. 
Experiments involve sampling from this distribution.

\begin{acknowledgments}
\emph{Acknowledgements}---
{I would like to thank Miles Blencowe, Arthur Fine, Lucien Hardy, Matt Leifer, and 
Lee Smolin for comments on earlier drafts.}
\end{acknowledgments}

\bibliography{BoltzEPR5} 
\bibliographystyle{apsrev4-1}

\end{document}